\documentclass[aip,apl,preprint]{revtex4-1} 

\usepackage{graphicx}
\usepackage{textcomp}
\usepackage{hyperref}

\draft 

\begin{document}

\title{Optical sensing with Anderson-localised light} 

\author{Oliver Joe Trojak}

\author{Tom Crane}

\author{Luca Sapienza}
\email{l.sapienza@soton.ac.uk}\homepage{www.quantum.soton.ac.uk}
\affiliation{Department of Physics and Astronomy, University of Southampton, Southampton, SO17 1BJ, United Kingdom}



\date{\today}


\begin{abstract}

We show that fabrication imperfections in silicon nitride photonic crystal waveguides can be used as a resource to efficiently confine light in the Anderson-localised regime and add functionalities to photonic devices. Our results prove that disorder-induced localisation of light can be utilised to realise an alternative class of high-quality optical sensors operating at room temperature. We measure wavelength shifts of optical resonances as large as 15.2\,nm, more than 100 times the spectral linewidth of 0.15\,nm, for a refractive index change of about 0.38. By studying the temperature dependence of the optical properties of the system, we report wavelength shifts of up to about 2\,nm and increases of more than a factor 2 in the quality factor of the cavity resonances, when going from room to cryogenic temperatures. Such a device can allow simultaneous sensing of both local contaminants and temperature variations, monitored by tens of optical resonances spontaneously appearing along a single photonic crystal waveguide. Our findings demonstrate the potential of Anderson-localised light in photonic crystals for scalable and efficient optical sensors operating in the visible and near-infrared range of wavelengths.

\end{abstract}


\maketitle 

Photonic crystals are optical devices based on a periodic modulation of the refractive index that allows the creation of photonic bandgaps where light propagation for certain wavelength ranges is forbidden \cite{PhC}. Bandgap engineering can allow the trapping of light in optical cavities or the confinement and guidance of light in the plane. Thanks to the high quality of the confinement achievable \cite{9million}, photonic crystals have been widely implemented for cavity quantum electrodynamics with solid-state emitters \cite{QED}, where the spontaneous emission dynamics of an emitter coupled to a photonic crystal cavity can be strongly modified \cite{weak} and even made reversible \cite{strong}. 
Photonic crystals can also be used as high-quality sensors \cite{sensing} since the presence of a contaminant can perturb the system by changing the local refractive index, resulting in variations of the wavelength of the light confined within the nanophonic device \cite{PhC_sensors}. Given the high quality of the resonances achievable, sub-nanometer shifts in the resonant wavelengths can be observed, thus allowing the detection of minimal refractive index changes. From this respect, photonic crystals surpass, for instance, plasmonic sensors, whose resonances can be as broad as 100\,nm \cite{plasmonic_sensor}, thus resulting in lower sensitivities. However, the main limitation in the development of photonic crystal cavities comes from the high level of accuracy required in the fabrication of devices, since even nanoscale imperfections can be detrimental to device performance \cite{Noda}. This has severely limited the scalability of devices based on photonic crystals, despite them being one of the most sensitive optical sensors that have been developed to date. 


\begin{figure}[htbp!]
\centering
\includegraphics[width=0.7\linewidth]{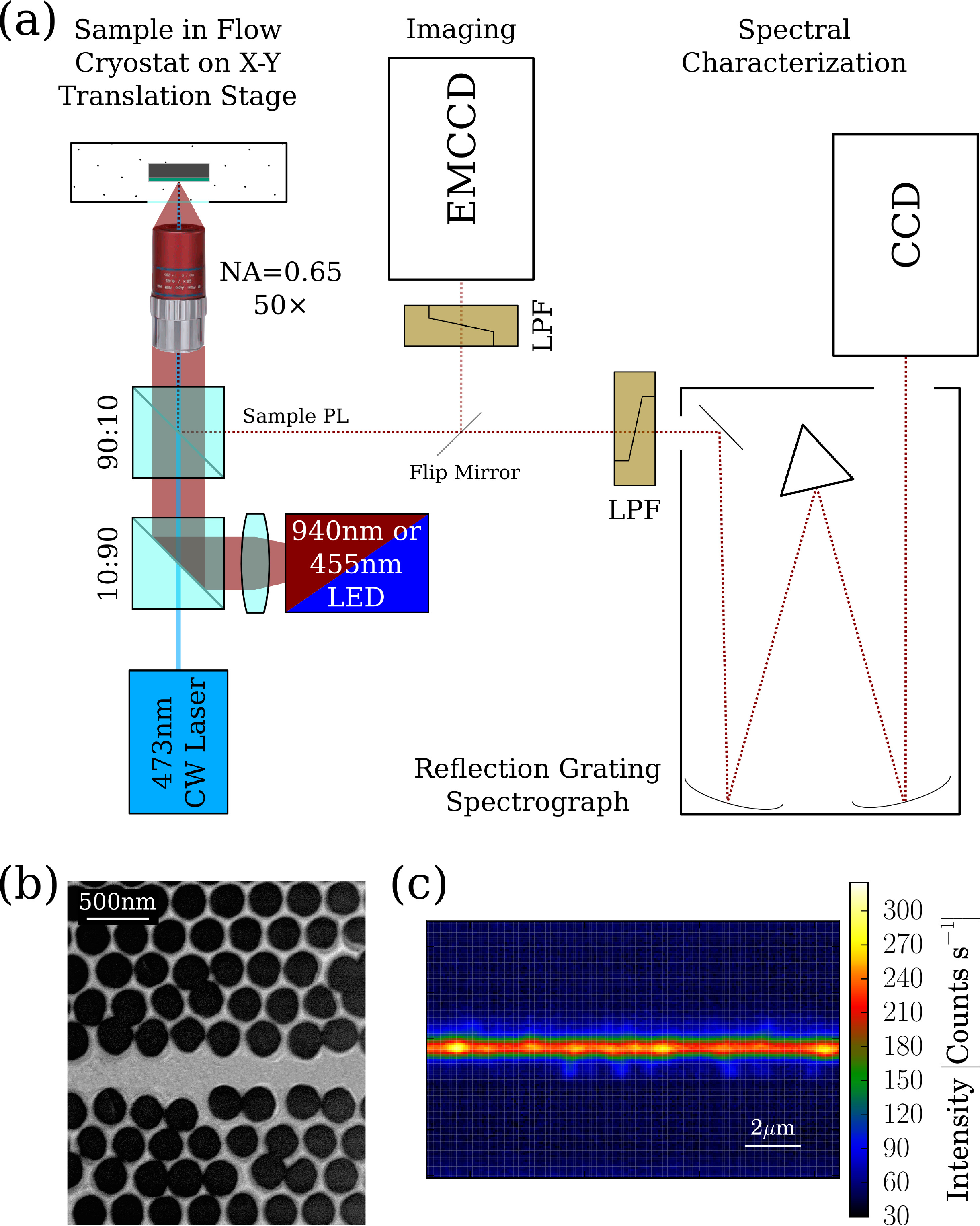}
\caption{Schematic of the confocal micro-photoluminescence (micro-PL) set-up (not to scale), comprising a continuous wave (CW) laser emitting at 473\,nm as excitation source and a 940\,nm light emitting diode (LED) for illumination, focused by a microscope objective (with numerical aperture NA=0.65) onto a sample, like the photonic crystal waveguide shown in the scanning electron microscope image in panel (b), placed on an X-Y translation stage. The 940\,nm LED is replaced by a 455\,nm LED for photoluminescence imaging. The imaging is carried out using an electron multiplied charge-coupled device (EMCCD) and the detection by using a CCD at the exit port of a reflection grating spectrometer used for spectral characterisation. LPF: Long Pass Filter. (c) Example of a photoluminescence image of the disorder-induced optical modes, probed by the silicon nitride luminescence, that appear as bright spots along the photonic crystal waveguide when excited by the 455\,nm LED, at room temperature.}
\end{figure}

To overcome this issue, we propose to follow a different approach, based on multiple scattering of light on imperfections as a means to achieve high-quality light confinement in the Anderson-localised regime \cite{us}. Here, we show that we can make use of fabrication imperfections as a means to add functionalities to the fabricated devices.  We report on photonic crystal optical sensors based on disorder-induced light confinement in photonic crystal waveguides in silicon nitride. We prove their suitability for the detection of liquid contaminants at room temperature and model their response to refractive index changes. Furthermore, we show that temperature can be used to tune and modify the quality factor of the cavity resonances, thus allowing local temperature sensing. Compared to engineered photonic crystal cavities, making use of disorder as a resource allows the spontaneous formation of tens of high-quality optical cavities in a fabricated device that does not require time-consuming optimizations or exact repeatability of the fabrication process - an important result in view of scalability of these photonic devices.

The samples are composed of a 250\,nm-thick silicon nitride layer deposited on a silicon substrate via plasma-enhanced chemical vapor deposition, using a combination of SiH$_{4}$ and NH$_{3}$ gases at a temperature of 350$^{\circ}$C and a pressure of 750\,mTorr. By means of electron-beam lithography, we write the photonic crystal pattern that is transfered onto the silicon nitride layer via an inductively coupled plasma reactive ion etch, based on SF$_{6}$ and C$_{4}$F$_{8}$. A KOH wet etch is used to undercut the silicon nitride, creating a free-standing photonic crystal membrane (see Fig.\,1b). By using finite-difference time-domain simulations, we optimise the photonic crystal parameters to confine light in the visible range of wavelengths (nominal values for the lattice constant and hole radius are 310\,nm and 110\,nm, respectively).
In such a device, the photonic crystal structure is based on the periodic change of the refractive index $n$ between silicon nitride ($n\sim$2) and air ($n$=1).
We fabricate photonic crystal waveguides 100\,$\mu$m long that show Anderson localisation of light due to multiple scattering on imperfections \cite{us}. These include for instance: deviations from the circularity and periodic position of the etched air holes, and imperfect verticality of the sidewalls \cite{Topolancik, Savona}. Such defects are inevitably introduced in the fabrication of the devices (intrinsic disorder) or can be engineered for instance by randomising the position of the air holes in the rows close to the waveguide channel (see Fig.\,1b). Multiple scattering of light on imperfections is responsible for trapping light in space within optical cavities that appear randomly distributed along the photonic crystal waveguide. The properties of disorder-induced light confinement in these devices, as a function of degree of disorder (from intrinsic to engineered), is discussed in detail in Ref.\cite{us}. However, we would like to stress that unavoidable fabrication imperfections are enough to achieve high-quality light localisation, therefore, no engineering of the disorder in the photonic crystal waveguide is needed.

\begin{figure}
\centering
\includegraphics[width=0.9\linewidth]{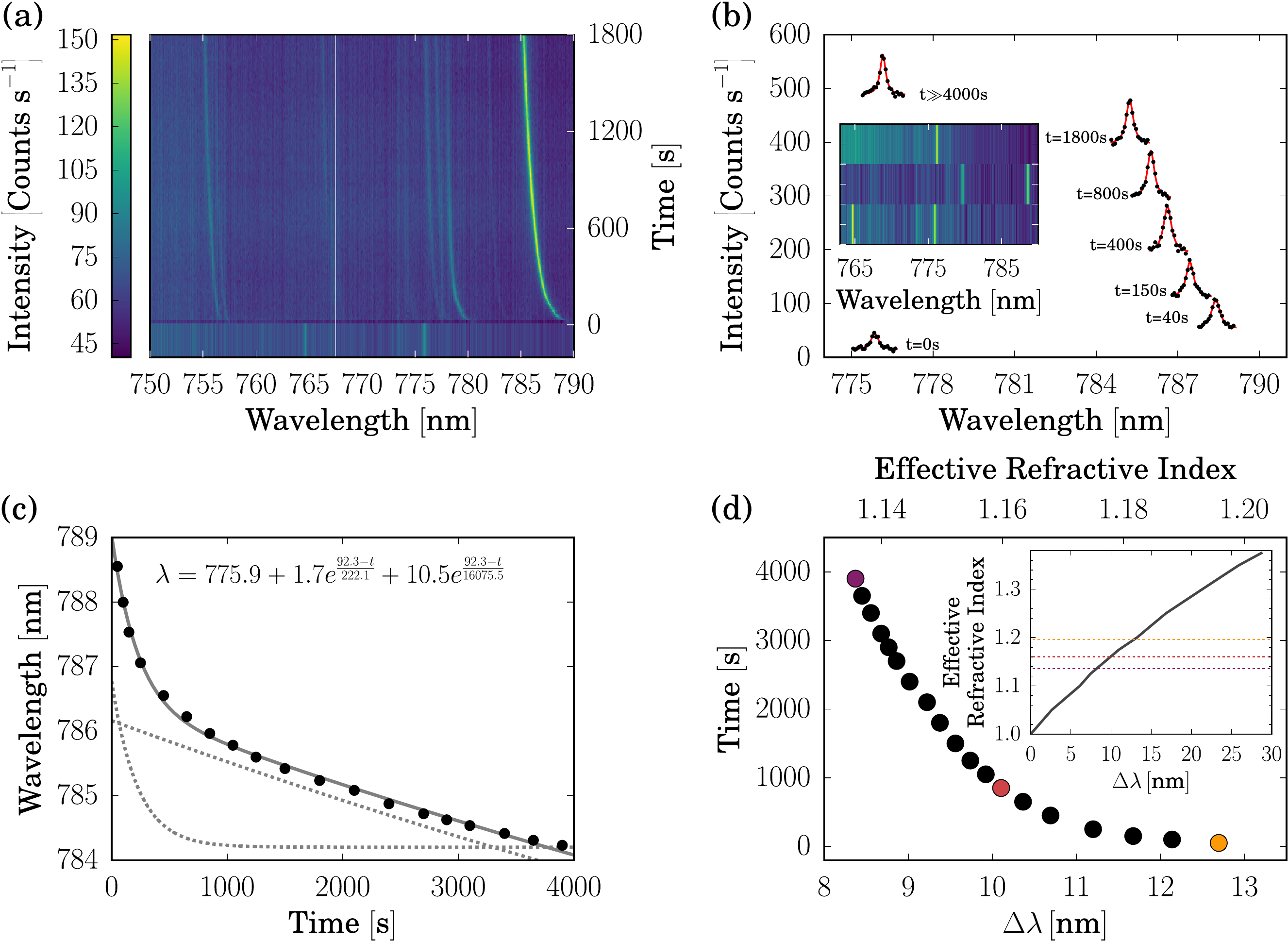}
\caption{(a) Photoluminescence spectra, collected under a laser excitation power density of 28\,kW/cm$^2$, at room temperature, plotted before isopropyl alcohol deposition (t=0, lower panel) and as a function of time after isopropyl alcohol deposition, showing the wavelength tuning of the optical resonance. (b) Example of spectra (symbols), collected under the same conditions as the spectra shown in panel (a), and their Lorentzian fits (solid lines) collected before isopropyl alcohol is applied to the photonic crystal waveguide (t=0) and as a function of time after isopropyl alcohol deposition. The spectra have been offset along the $y$-axis for clarity.  Inset: Zoom in of the spectra collected before isopropyl alcohol deposition (lower panel), 40\,s after deposition (middle panel) and $\gg$4000\,s after deposition, proving reversibility. (c) Wavelength of the optical resonances plotted as a function of time, as obtained from the Lorentzian fits (symbols, the error bars are too small to be visible). A double-exponential fit to the data (solid line), shows the time evolution of the process. The dashed lines represent the single exponential components of the decay curve, offset for clarity. (d) Experimental wavelength shifts $\Delta\lambda$ of cavity resonances compared to finite-difference time-domain simulations for different effective refractive indices (inset) used for the filling of the photonic crystal holes (see main text for more details). The coloured dots correspond to the values of the effective refractive indices highlighted by the corresponding dashed lines in the inset.}
\end{figure}

Light localisation is probed by means of micro-photoluminescence spectroscopy, using the confocal set-up presented in Fig.\,1a. The sample is illuminated by a 940\,nm light emitting diode (LED) for sample imaging or a 455\,nm LED for photoluminescence imaging via an electron multiplied charge-coupled device (EMCCD). An example of a photoluminescence image is shown in Fig.\,1c where each bright spot corresponds to a different disorder-induced localised optical mode \cite{us}. The intrinsic silicon nitride luminescence \cite{intrinsic_PL} is locally excited with a laser emitting at 473\,nm (with an excitation spot of $\sim$2\,$\mu$m in diameter) and the photoluminescence signal is sent to a grating spectrometer, equipped with a silicon charge-coupled device (CCD), for spectral analysis. When scanning the excitation laser along a waveguide, a large number of optical cavities appear as sharp spectral resonances. Their quality factors, evaluated as the central wavelength divided by the full width half maximum of the spectral peak, show a distribution ranging from one to ten thousand \cite{us}. It is worth noting that such disorder-induced optical cavities show quality factors that exceed those of engineered two-dimensional silicon nitride photonic crystal cavities confining light at comparable wavelenghts \cite{Q, Q1, Benson}. Compared to engineered photonic crystal cavities, the formation of a large number of high-quality optical cavities within the same device is particularly suitable for optical sensing, since many optical resonances can be monitored simultaneously. Making use of fabrication imperfections, rather than highly engineering photonic devices that require nanoscale-accurate fabrication procedures \cite{Noda} and in some cases several iterations, is an asset for increasing the scalability of the fabricated devices.

We have tested the sensitivity of our device when a small amount (estimated to be $\sim$20\,p$\ell$ on the photonic crystal waveguide area of approximately 10$^3$\,$\mu$m$^2$) of a contaminant, isopropyl alcohol (whose refractive index is 1.38), is deposited on the surface of a photonic crystal waveguide. Based on the quality factors that we achieve, our devices are already competitive with current photonic crystal structures specifically engineered for sensing applications \cite{7500Q, PhC_sensors}. Figure 2a shows how several disorder-induced optical cavities are shifted by the presence of the contaminant. In Fig.\,2b, we show an example of a disorder-induced resonance appearing at 775.9\,nm shifted by the presence of the alcohol to 788.5\,nm: this corresponds to a wavelength shift of 12.6\,nm, more than 30 times the linewidth of 0.4\,nm of the spectral resonance. The resonance at 764.4\,nm shown in the inset is  0.15\,nm broad and shifts of 15.2\,nm, more than 100 times its linewidth. These results prove the sensitivity of our system to small amounts of contaminants, for a relatively small refractive index change. Our devices show shift/linewidth ratios of the optical resonances that even exceed values reported for state-of-the-art optimised sensors based on photonic crystal cavities \cite{sensor2}. Given the volatility of the alcohol used at ambient temperature and pressure, the shift of the spectral position of the optical resonance varies as a function of time after deposition (see Fig.\,2a, 2b). It is worth noting that the wavelength shift is characterised by a double exponential decay (see Fig.\,2c): this can be explained by the fact that the alcohol will evaporate at a faster rate from the top of the waveguide channel and at a slower rate when confined within the photonic crystal holes. As shown in Fig.\,2a and 2b, given the brightness of the spectral resonances in our device, the sensing process can be monitored in real time and proves to be reversible (see inset of Fig.\,2b), following the time dependence described by the equation shown in Fig.\,2c.  
In order to model the shifts in the optical resonances that we observe, we have carried out finite-difference time-domain simulations that allow us to evaluate the wavelengths of photonic crystal optical resonances when changing the refractive index of the holes. When no contaminants are present, the holes are considered to be filled by air, with refractive index of 1. Holes completely filled by isopropyl alcohol would have a refractive index of 1.38. When the holes are partially filled, we consider an effective refractive index whose value reflects the amount of air and alcohol filling the holes. As shown in Fig.\,2d, by modifying the refractive index of the holes between 1 and 1.38, we are able to reproduce the shifts in the cavity resonances that we observe. Our simulations suggest that the holes are not fully filled by isopropyl alcohol in the experiments that we carry out, suggesting that an even larger wavelength shift, up to 30\,nm, could be potentially achieved in the system under study.
Our simulations also prove that different amounts of a contaminant, and/or contaminants with different refractive indices, provide different wavelength shifts of the disorder-induced optical resonances: a calibration of the system would thus allow not only the verification of the presence of a contaminant, but also the evaluation of its quantity and/or its refractive index.

\begin{figure}[t!]
\centering
\includegraphics[width=0.9\linewidth]{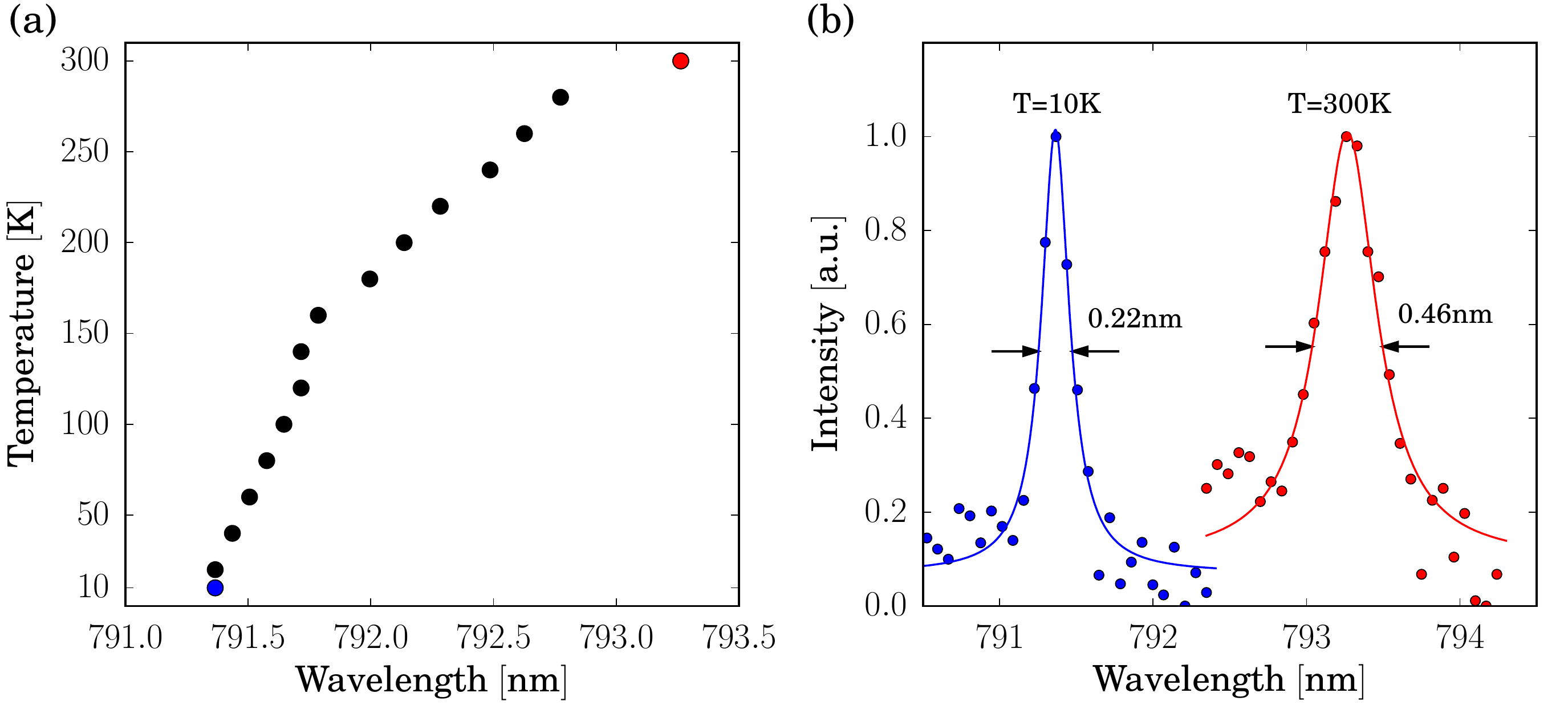}
\caption{(a) Central wavelength of the cavity peaks, extracted by the Lorentzian fits of the photoluminescence spectra, collected under a laser excitation power density of 28\,kW/cm$^2$, as a function of the sample temperature, varied with steps of 20\,K (the error bars are too small to be visible). (b) Normalised photoluminescence spectra collected at a temperature of 10 (blue dots) and 300\,K (red dots) and their Lorentzian fits (solid lines). The full width half maximum of the resonances is also shown.}
\end{figure}

Furthermore, we have tested the temperature dependence of the optical resonances. The sample is placed in an liquid-Helium flow-cryostat where the sample temperature, measured by a three point calibrated rhodium iron temperature sensor placed in the cold finger where the sample is mounted, can be varied from 300 to 10\,K. Photoluminescence spectra are collected while the temperature is varied in defined steps. When varying the sample temperature, the refractive index of silicon nitride is modified and, given the sensitivity of photonic crystal structures to small refractive index variations, this results in a spectral tuning of the optical resonances. As shown in Fig.\,3a, we observe a blue-shift of the spectral peak of up to $\sim$2\,nm, when changing the temperature from 300 to 10\,K. Such a shift is fully reversible and provides a local probe of the sample temperature in correspondence to the specific optical mode under examination. 

The temperature tuning of the optical resonances can be used for tuning an optical mode into and out of resonance, for instance, with the emission line of defect centers in diamond \cite{diamond} and two-dimensional materials \cite{2D} that emit in the visible and near-infrared range of wavelengths, thus allowing to carry out cavity quantum electrodynamics experiments \cite{QED}. Furthermore, we observe a variation of the quality factor of the optical resonance that increases of about a factor 2 when going from 300 to 10\,K (see Fig.\,3b). This can be explained by reductions in cavity losses \cite{Q_T}, showing that the light-matter interaction can be further enhanced at cryogenic temperatures.

In conclusion, we have demonstrated optical sensing with disorder-induced optical cavities in photonic crystal waveguides in the Anderson-localised regime. We have observed reversible spectral shifts up to 100 times the linewidth of the spectral resonances for liquid contaminants providing a refractive index change of $\sim$0.38. Furthermore, we have shown temperature shifts of up to $\sim$2\,nm when varying the temperature from 300 to 10\,K, accompanied by a cavity quality factor increase of a factor 2. These results show that disorder-induced light confinement in silicon nitride photonic crystals is suitable for the development of high sensitivity, scalable, room temperature optical sensors and as a platform for cavity quantum electrodynamics experiments. Such experiments take advantage of the spontaneous formation of tens of high-quality optical cavities along the fabricated photonic crystal waveguides, allowing simultaneous sensing with different optical resonances, with no need for multiple iterations of the fabrication process. 
For the application of our devices in single particle sensing \cite{particle_sensing}, the deposition of single molecules on the sensors could be carried out by means of nanofluidics \cite{nanofluidics}. The ability of locating optical modes with nanometer-scale accuracy in our system, thanks to the photoluminescence imaging technique that we have developed (see Fig. 1c), will allow a controlled deposition of single molecules, that could, for instance, be functionalised in order to stick to the surface in correspondence to specific optical cavities.
The implementation of disorder imperfections to confine light relaxes the stringent requirements of nanoscale accuracy in the fabrication of the devices, that could also be realised using techniques less costly than electron-beam lithography like deep ultra-violet photolithography \cite{UV} and nano-imprint lithography \cite{nanoimprint}, desirable for larger scale production.

\section*{Acknowledgments}

We would like to thank Hector Stanyer and Josh Nevin for their early contribution to this work and the Annual Adventures in Research award scheme for partial financial support.


\begin{thebibliography}{50}



\bibitem{PhC} J.D. Joannopoulos, S.G. Johnson, J.N. Winn, R.D. Meade, \textit{Photonic crystals: molding the flow of light}, Princeton University Press (2011).

\bibitem{9million} H. Sekoguchi, Y. Takahashi, T. Asano, S. Noda, Optics Express \textbf{22}, 916 (2014).

\bibitem{QED} M. Pelton, Nature Photonics \textbf{9}, 427 (2015).

\bibitem{weak} A. Kress, F. Hofbauer, N. Reinelt, M. Kaniber, H. J. Krenner, R. Meyer, G. Böhm, and J. J. Finley, Physical Review B \textbf{71}, 241304(R) (2005).

\bibitem{strong} K. Hennessy, A. Badolato, M. Winger, D. Gerace, M. Atat\"ure, S. Gulde, S. Fält, E. L. Hu, A. Imamo\u glu, Nature \textbf{445}, 896 (2007).

\bibitem{sensing} J. Hodgkinson, R.P. Tatam, Measurement Science and Technology \textbf{24}, 012004 (2013).

\bibitem{PhC_sensors} Y. Zhanga, Y. Zhaoa, R-Q. Lv, Sensors and Actuators A \textbf{233}, 374 (2015).

\bibitem{plasmonic_sensor} M. Li, S.K. Cushing, N. Wu, Analyst \textbf{140}, 386 (2015).

\bibitem{Noda} Y. Akahane, T. Asano, B.-S. Song, and S. Noda, Nature, \textbf{425}, 944 (2003).

\bibitem{us} T. Crane, O.J. Trojak, J.P. Vasco, S. Hughes, L. Sapienza, ACS Photonics \textbf{4}, 2274 (2017).

\bibitem{Topolancik} J. Topolancik, B. Ilic, F. Vollmer, Physical Review Letters \textbf{99}, 253901 (2007).

\bibitem{Savona} V. Savona, Physical Review B \textbf{83}, 085301 (2011).

\bibitem{intrinsic_PL} J. Kistner, X. Chen, Y. Weng, H.P. Strunk, M.B. Schubert, J.H. Werner, Journal of Applied Physics \textbf{110}, 023520 (2011).

\bibitem{Q} M. Makarova, J. Vu\v{c}kovi\'c, H. Sanda, Y. Nishi, Applied Physics Letters \textbf{89}, 221101 (2006). 

\bibitem{Q1} M.M. Murshidy, A.M. Adawi, P.W. Fry, D.M. Whittaker, D.G. Lidzey,, Journal of the Optical Society of America B \textbf{27}, 215 (2010).

\bibitem{Benson} M. Barth, N. N\"usse, J. Stingl, B. L\"ochel, O. Benson, Applied Physics Letters \textbf{93}, 021112 (2008).

\bibitem{7500Q} S.H. Mirsadeghi, E. Schelew, J.F. Young, Applied Physics Letters \textbf{102}, 131115 (2013).


\bibitem{sensor2} E. Chow, A. Grot, L.W. Mirkarimi, M. Sigalas, G. Girolami, Optics Letters \textbf{29}, 1093 (2004).

\bibitem{diamond} I. Aharonovich, A.D. Greentree, S. Prawer, Nature Photonics \textbf{5}, 397 (2011).


\bibitem{2D} F. Xia, H. Wang, D. Xiao, M. Dubey, A. Ramasubramaniam, Nature Photonics \textbf{8}, 899 (2014).

\bibitem{Q_T} Y. Gong, M. Makarova, S. Yerci, R. Li, M.J. Stevens, B. Baek, S. Woo Nam, R.H. Hadfield, S.N. Dorenbos, V. Zwiller et al., Optics Express \textbf{18}, 2601 (2010).

\bibitem{particle_sensing} S.H. Choi and Y.L. Kim, Applied Physics Letters \textbf{100}, 041101 (2012).

\bibitem{nanofluidics} L. Bocquet and E. Charlaix, Chemical Society Reviews \textbf{39}, 1073 (2009).

\bibitem{UV} Y. Ooka, T. Tetsumoto, A. Fushimi, W. Yoshiki, T. Tanabe, Scientific Reports \textbf{5}, 11312 (2015).

\bibitem{nanoimprint} T. Senn, J. Bischoff, N. N\"usse, M. Schoengen, B.L\"ochel, Photonics and Nanostructures - Fundamentals and Applications \textbf{9}, 248 (2011).



\end{thebibliography}
\end{document}